\documentstyle[11pt,paspconf,epsf]{article}

\begin{document}

\title{The Evolution of Radio-Loud Quasars at High Redshift}

\author{I. M. Hook, P. A. Shaver}
\affil{European Southern Observatory, Karl Schwarzschild Strasse-2,
D85748 Garching b. M\"{u}nchen, Germany}

\author{R. G. McMahon}\affil{Institute of Astronomy,
Madingley Road, Cambridge, U.K.}

\begin{abstract}
We present results from two surveys for radio-loud quasars. These are
(1) a northern survey reaching $\rm S_{5GHz}=0.2$Jy and (2) the Parkes
QSO survey reaching $\rm S_{2.7GHz}=0.25$Jy, which together cover most
of the sky. Both surveys use digitised plate material to identify the
majority of the radio sources and CCD images to identify the optically
fainter sources. The northern survey was targeted specifically at
$z>3$ QSOs and has produced more than 20, one of which, at $z=4.72$,
is the most distant radio source known.  This sample is identified to
the POSS-I plate limit (20 mag) over an area of 3.7sr, or 12 000 sq
deg. With the addition of CCD identifications, an effective sub area
of $2400$ sq degrees is 97\% identified. In the South the Parkes QSO
survey now has complete redshift information for all 442 flat-spectrum
quasars.  These cover the redshift range $z\sim 0$ to $z\sim 4.5$ and
show a clear drop-off in space density at redshifts above 3. The form
of this decline is remarkably similar in form to that seen in
optically-selected samples of quasars. Since radio emission is
unaffected by dust this implies that dust has a minimal effect on the
observed drop-off seen in optical samples. Analysis of our
radio-selected samples suggests that the decline in space density at
high redshift is more pronounced for less powerful radio sources. We
are planning a new survey that will be sensitive to bright quasars
with redshifts up to 6.
\end{abstract}


\section{Introduction}

The mere existence of QSOs and their host galaxies at high redshift
places interesting constraints on theories of large-scale structure
and galaxy formation (Haehnelt \& Rees 1993, Efstathiou \& Rees
1988). However, recent optical surveys have produced conflicting
results for the QSO space density at high redshift, due to the
complicated selection effects introduced when picking out
high-redshift QSOs from more numerous galactic stars with similar
optical colours. {\it Radio}-based selection methods have the
advantage of simpler selection criteria. Also, since radio emission is
unaffected by dust, the intrinsic evolution of the QSO population can
be observed, unaffected by obscuration.  QSOs found in radio-based
surveys provide an unbiased (by the effects of dust) sample of damped
Ly-$\alpha$ absorption systems that can be used to study chemical
abundances to the highest observable redshifts.

\section{Two surveys for radio-loud quasars}

\begin{figure}
\plotfiddle{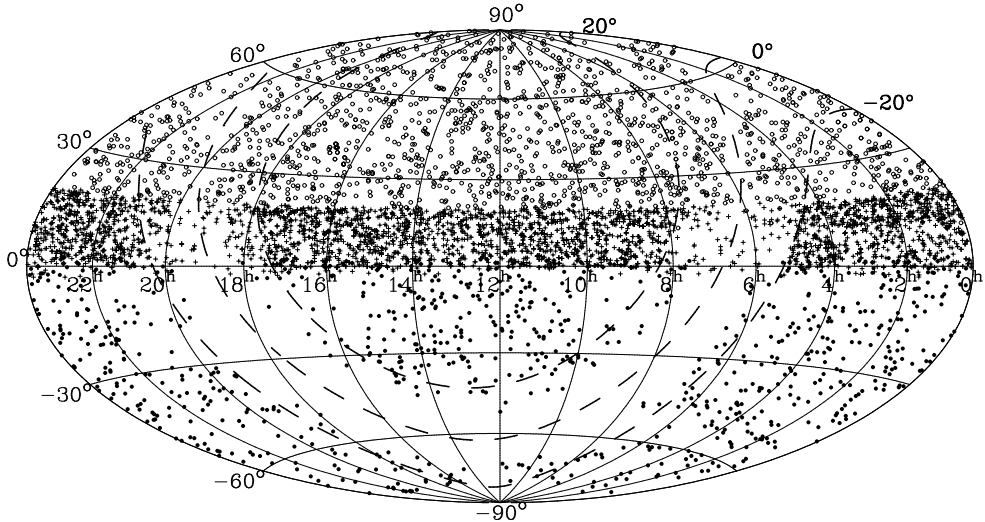}{6.0cm}{0.0}{80.0}{80.0}{-260.0}{-230.0}
\caption{Sky coverage of the three radio-samples that form the basis
of our surveys. These are the 0.2Jy flat-spectrum sample of Patnaik et
al (1992) for $\delta > 20^{\circ}$, the 0.1Jy MG-VLA sample (Lawrence
et al 1986) in the region $0^{\circ} < \delta < 20^\circ$ and
flat-spectrum sources from the Parkes catalogue in the South. Almost
the whole sky (excluding the galactic plane, shown by the dashed
lines) is covered to a depth of 0.25Jy or fainter. Only regions that
are complete to a well-defined flux limit are used for the analysis of
quasar evolution.}
\label{sky}
\end{figure}

Since high-redshift quasars are rare, surveys to reach $z>3$ must
cover a large area. We are carrying out two surveys using three radio
samples that cover most of the sky to a limit of 0.25Jy or fainter
(see Figure~\ref{sky} and caption). The northern survey uses the 0.2Jy
flat-spectrum sample of Patnaik et al (1992) and the 0.1Jy MG-VLA
sample (Lawrence et al 1986), which together contain more than 4400
sources. The southern survey uses the revised Parkes catalogue (Wright
\& Otrupcek, 1990). Since these samples are defined at high frequency
(5GHz for the northern samples and 2.7GHz for the Parkes sample), they
contain a high fraction of core-dominated flat-spectrum sources,
usually identified with quasars.  Moreover, spectral index information
is avaliable for all the sources, allowing us to select flat-spectrum
objects.  In addition the sources have accurate positions from the VLA
and/or the Australia telescope, which are necessary for making
unambiguous optical identifications.

The first stage of optical identification is carried out using
digitised sky-survey plates. CCD images are then obtained for the
fainter sources. Since the northern survey was aimed at finding
high-redshift quasars, only the optically red, stellar identifications
were followed up spectroscopically. For the Parkes sample redshifts
were obtained for all the stellar identifications. Details of the two
surveys are given below.

\subsection{The Northern High-redshift Radio-loud QSO Sample.} 

\begin{figure}
\plotfiddle{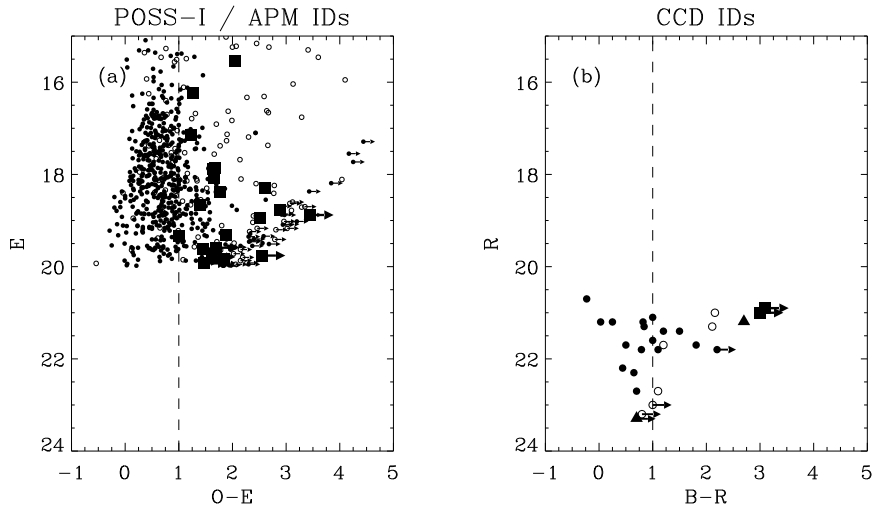}{7.0cm}{0.0}{70.0}{70.0}{-200.0}{-210.0}
\caption{Colour-magnitude diagrams for flat-spectrum sources with $\rm
S\ge 0.2$Jy from the northern survey. (a) APM/POSS-I identifications
covering an area of 12 000 sq deg. (b) CCD identifications for a 2400
sq deg sub-area. Stellar sources are shown as filled circles and open
circles represent galaxies. The $z>3$ quasars are shown as
squares. The two triangles represent red, stellar objects that do not
yet have a spectrum. The dashed lines show the colour limits used to
define the spectroscopic sample.}
\label{colmag}
\end{figure}

POSS-I plates in the E and O bands, scanned by the APM (Automated
Plate Measuring facility, Cambridge U.K.), were used to identify
flat-spectrum ($\alpha \ge -0.5$, $\rm S\propto \nu^\alpha$) sources
with $\rm S_{5GHz}\ge 0.2$Jy to the plate limit of $\rm E = 20$mag
over an area of 3.7sr (12 000 sq deg).  Their colour magnitude diagram
is shown in Figure~\ref{colmag}a. Since high-redshift quasars have
redder optical colours than their low-redshift counterparts due to
absorption by intervening Ly-$\alpha$, only red ($\rm O-E\ge 1.0$),
stellar identifications were followed up spectroscopically. This
resulted in the discovery of 20 $z>3$ quasars (Hook 1994, Hook et al
1995, 1996). Another 5 $z>3$ QSOs with radio fluxes $\rm 0.1Jy \le
S_{5GHz} \le 0.2Jy$ were found in the region covered by the MG-VLA
sample.

\begin{figure}[t]
\plotfiddle{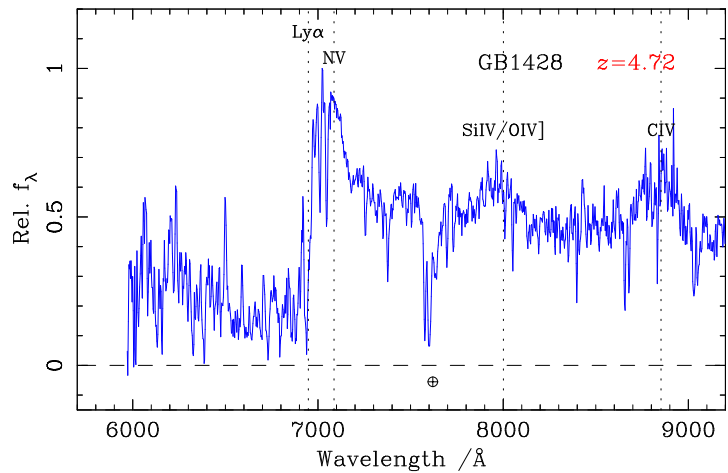}{5.0cm}{0.0}{50.0}{50.0}{-150.0}{-210.0}
\caption{The optical spectrum of GB1428, the most distant known radio
source and the third highest redshift QSO known (Hook \& McMahon
1998). This object was discovered among CCD identifications of
flat-spectrum radio sources with $\rm S\ge 0.2Jy$.  The spectrum shows
unusually weak emission lines for a radio-loud quasar, consistent with
the scenario that it is beamed (Fabian et al 1997). Note that the
spectrum has not been corrected for the sky absorption feature at
$\sim 7600$\AA.}
\label{spec}
\end{figure}

About 16\% of the radio sources remained unidentified after this stage
and B and R-band CCD identifications are being obtained for these. So
far $\sim 40$ have been observed and a colour-magnitude diagram for
those detected in the R-band is shown in
Figure~\ref{colmag}b. Combining the POSS-I and CCD identifications
plus observations of these radio sources from the literature
(including the identification of one source as a QSO with $z=3.82$,
Vermeulen et al 1996), an effective area of 2400 sq deg is now 97\%
identified.  10 objects are still not identified to a limit of $\rm
R\sim 23$mag. Two new $z>4$ QSOs were found during spectroscopy of the
10 reddest CCD identifications.  One of these, at $z=4.72$ is the
highest redshift radio source known (Hook \& McMahon 1998). Its
spectrum is shown in Figure~\ref{spec}. It is also the highest
redshift X-ray source known and is probably strongly beamed (Fabian et
al 1997).

\begin{figure}[h]
\plotfiddle{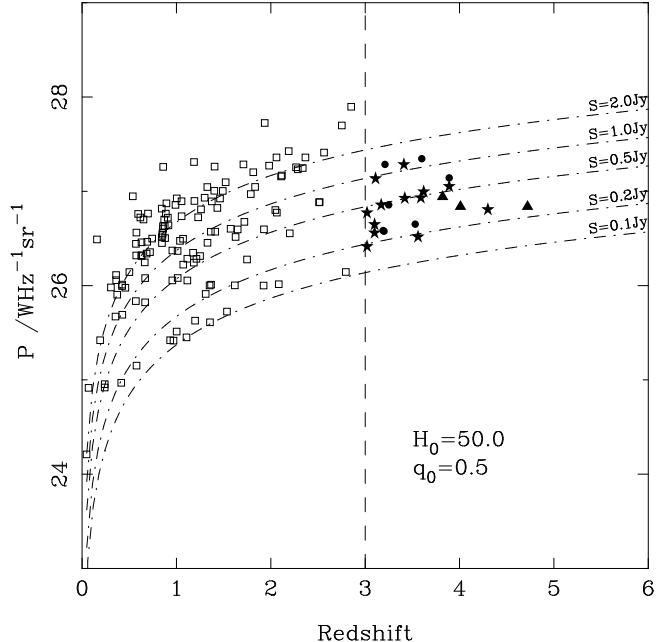}{7.0cm}{0.0}{50.0}{50.0}{-170.0}{-60.0}
\caption{Distribution of $z>3$ radio-loud QSOs with $\rm S\ge0.2$Jy
from the northern sample (solid symbols) on the radio power vs
redshift plane, plotted with flat-spectrum QSOs with $z<3$ from Dunlop
\& Peacock 1990 (open squares). Circles are QSOs from the MG-VLA
sample; stars represent QSOs from the $\rm S\ge0.2Jy$ flat-spectrum
sample identified on APM/POSS-I scans and triangles are sources from
the same radio sample identified on CCD images.
}
\label{pz}
\end{figure}

\begin{figure}[h]
\plotfiddle{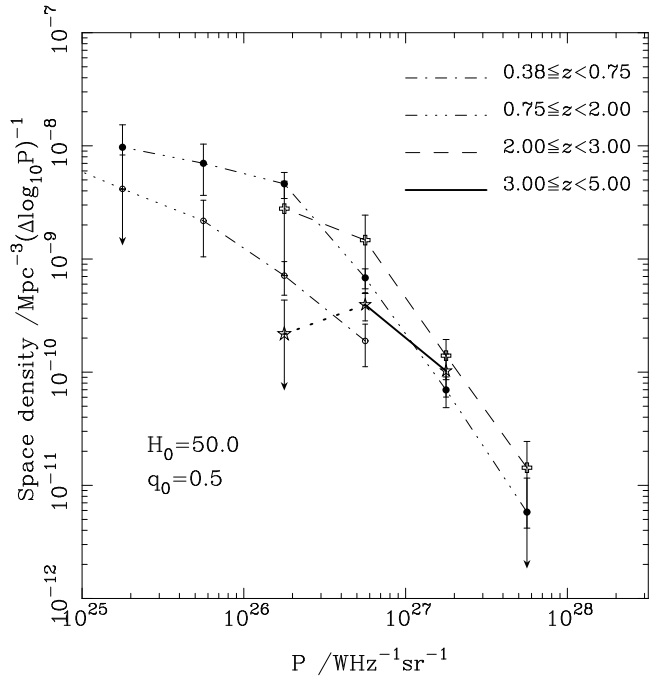}{7.0cm}{0.0}{50.0}{50.0}{-170.0}{-60.0}
\caption{The luminosity function of flat-spectrum QSOs as determined
from the data shown in Figure~\ref{pz}. Notice that the space density
at $z>3$ is lower than at $z\sim 2$, particularly for weaker sources.
Downward arrows indicate that the error bar extends to zero since
there is only one point in these bins.}
\label{lf}
\end{figure}

We are able to compute the luminosity function of radio-loud QSOs at
high redshift using the $z>3$ QSOs discovered among the POSS and CCD
identifications. The distribution of the QSOs on the radio-luminosity
versus redshift plane is shown in Figure~\ref{pz} along with
flat-spectrum QSOs at $z<3$ from Dunlop \& Peacock (1990). The binned
luminosity function for these two data sets is shown in
Figure~\ref{lf}. Notice that the luminosity function at $z>3$ is lower
than that at $z\sim 2$, particularly for sources of lower radio
luminosity.

\subsection{The Parkes QSO Sample.} 

\begin{figure}[h]
\plotfiddle{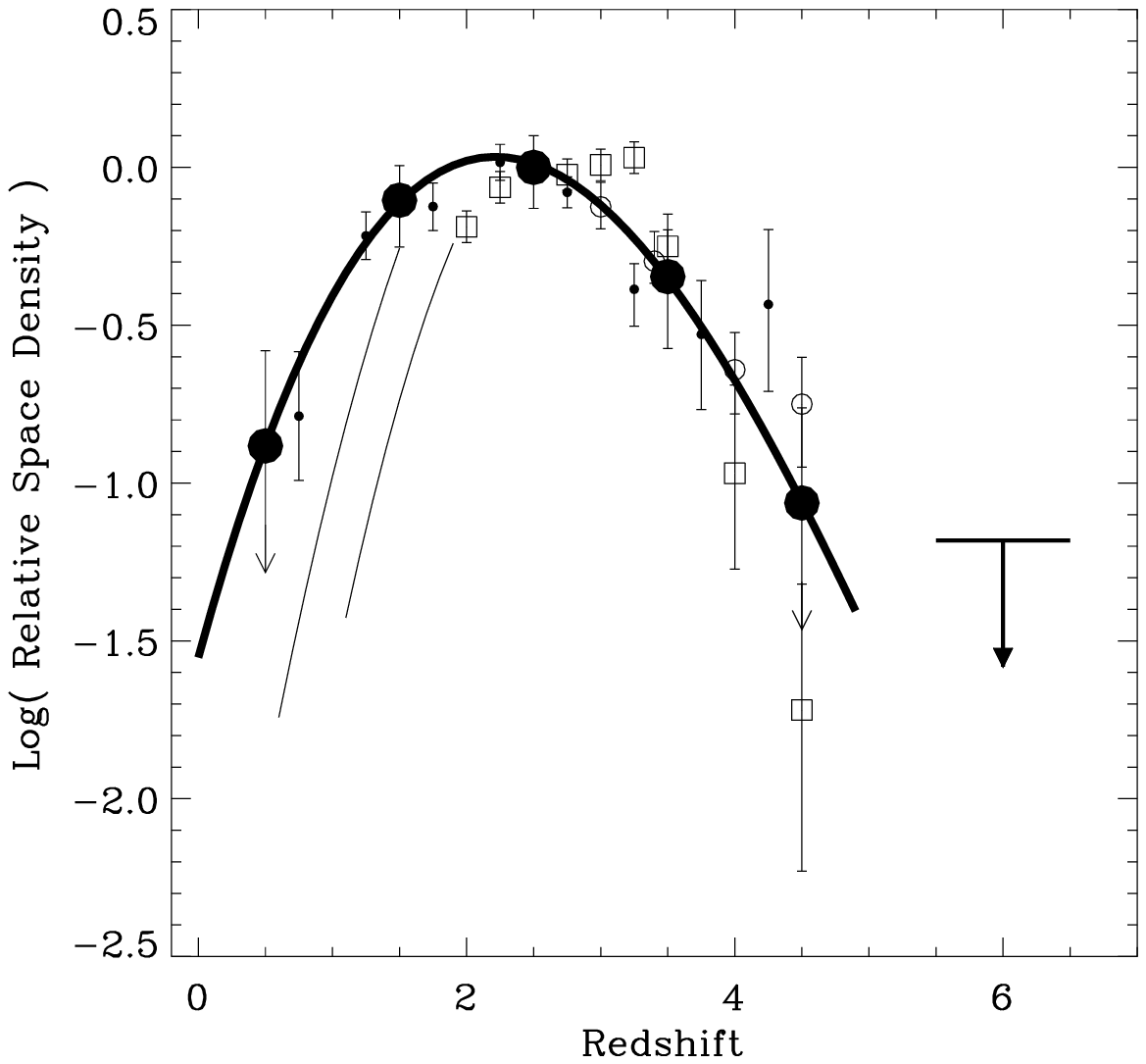}{7.0cm}{0.0}{75.0}{75.0}{-180.0}{-280.0}
\caption{Space densities, normalized to $z \sim 2-3$ and plotted as a
function of redshift, for the Parkes flat-spectrum radio-loud quasars
with $\rm P_{lim} > 7.2$ 10$^{26}$ W Hz$^{-1}$ sr$^{-1}$ ({\huge
$\bullet$}). The number of such objects found in the individual
redshift ranges are 1 (0$<$$z$$<$1), 12 (1$<$$z$$<$2), 15
(2$<$$z$$<$3), 6 (3$<$$z$$<$4), and 1 (4$<$$z$$<$5); the error bars
correspond to $\pm \sqrt{N}$. The thick curve is a cubic fit to these
data. The upper limit shown in the redshift range $5 < z < 7$ is taken
from Shaver et al (1996a). For comparison, similarly normalized space
densities are also shown for the optically-selected quasar samples of
Warren {\it et al.} (1994) ($\Box$), Schmidt {\it et al.} (1995)
($\circ$), and Hawkins \& V\'{e}ron (1996) ({\tiny $\bullet$}). The
thin lines represent luminosity functions from Boyle (1991) and Hewett
{\it et al.} (1993) used by Warren {\it et al.} and Schmidt {\it et
al.} respectively as lower-redshift continuations of their
high-redshift space densities.}
\label{drop}
\end{figure}

Initial identifications of flat-spectrum sources ($\alpha\ge -0.4$)
from the Parkes catalogue were made by Shaver et al (1996a) using
COSMOS scans of UKST plates to a limit of $\rm B_J \sim 22mag$. CCD
identifications were obtained for the remaining optically-fainter
sources. Since all the stellar sources have been identified in the B
band, Shaver et al (1996a) were able to conclude there are no $z>5$
QSOs in the sample (see the limit in Figure~\ref{drop}) and hence
there must be a drop-off in the space density of radio-loud quasars at
$z>3$.

The next stage of this study, namely to obtain redshifts for a
complete sub-sample of 442 stellar identifications (quasars) with $\rm
S_{2.7GHz}\ge 0.25$Jy, has recently been completed. The most distant QSO found
in this sample has $z=4.46$ (Shaver et al 1996b). Figure~\ref{drop}
shows a preliminary analysis of space density as a function of
redshift for objects with radio luminosity greater than $\rm
P_{lim}=7.2\times10^{26}WHz^{-1}sr^{-1}$, corresponding to the flux
limit of the sample at $z=5$ (for $\rm H_0=50 kms^{-1}Mpc^{-1}$ and
$q_0=0.5$).  The Figure shows that the decline in space density of
luminous Parkes QSOs at $z>3$ has a remarkably similar form to that
seen in optically-selected samples of QSOs. Since the radio sample is
unaffected by dust this suggests that dust has a minimal effect on the
observed drop-off in space density of optically-selected QSOs. This
also has implications for the effect of dust on recent measurements of
the star formation rate from high-redshift galaxies. We are working on
quantifying this result in the context of the obscuration models of
Fall \& Pei (1993).

\section{QSO Evolution \& A New Large Survey}
\begin{figure}[h]
\plotfiddle{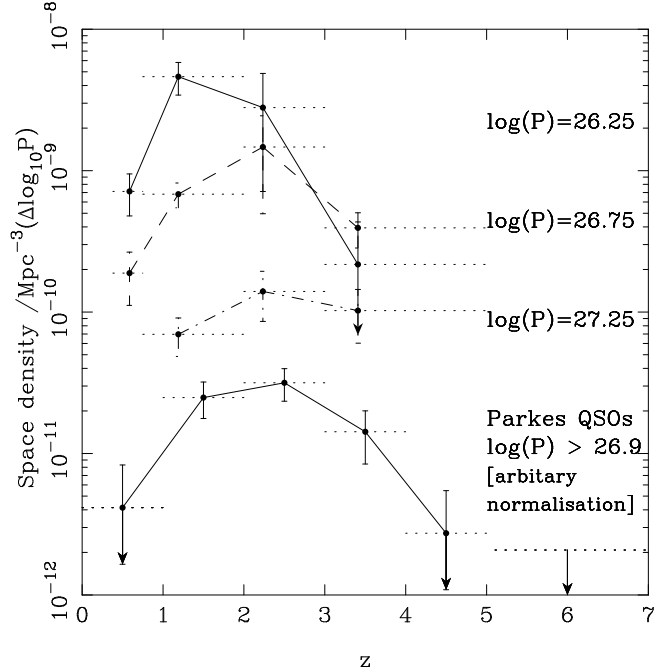}{7.0cm}{0.0}{50.0}{50.0}{-150.0}{-70.0}
\caption{Space density as a function of redshift for sources with
various radio luminosity P. The top three curves are from the northern
sample combined with data from Dunlop \& Peacock (1990) -see caption
of Figure~\ref{pz}.  The lower curve is from the Parkes sample (see
Fig.~\ref{drop}) and is plotted with arbitary normalisation. The
dotted lines show the extent of the redshift bin used to compute each
point. }
\label{rhoz}
\end{figure}

The results from the above surveys suggest that the peak in QSO space
density occurs at higher redshifts for stronger sources
(Fig~\ref{rhoz}). Thus a survey to reach higher redshifts should
concentrate on relatively bright objects and cover a large area,
rather than reaching very faint limits over a small region.

We are planning a very large survey for $z>4$ QSOs, making use of new
wide-area datasets.  The survey covers 6000 sq deg and is based on
flat-spectrum sources with $\rm S_{5GHz}\ge 50mJy$ from the 5GHz PMN
sample. The 1.4GHz NRAO-VLA Sky Survey (NVSS, Condon et al 1994),
which now covers the whole sky north of $-40^{\circ}$, will provide
spectral index information and the accurate positions needed to make
unambiguous optical identifications.  The UKST I-band survey of the
southern sky is now nearing completion and the plates are being
scanned at the APM. As in the northern survey described above,
candidate high-redshift QSOs will be selected based on positional
coincidence, red optical colour and `stellar' image classification.
By using I-band plates this survey is sensitive to bright QSOs out to
$z\sim 6$. Extrapolating our previous results to 50mJy, $8-24$
radio-loud objects with $z>4$ should be found ($2-6$ times the number
currently known) plus up to 2 with $z>5$.

\section{Conclusions}

There are 4 main conclusions from this work.

\begin{itemize}
\item Radio surveys are efficient for finding high-redshift
  QSOs. There are now many tens of $z>3$ radio-selected quasars known
  including GB1428 at $z=4.72$, the third highest redshift quasar
  known.
      
\item The space density of radio-selected QSOs shows a turnover at
       high-redshift ($z>3$). Since radio emission is unaffected by
       dust, this decline at high redshift is not an effect of
       obscuration.
\item The decline in space density seen at high redshift has a similar
form to that of optically-selected QSOs.
\item The two previous results lead to the conclusion that the effect
of dust on the observed QSO drop-off is minimal.

\end{itemize}

\end{document}